\documentclass[aps,pra,showpacs,amssymb,twocolumn]{revtex4-1}
\usepackage{graphicx}
\usepackage{subfig}
\usepackage{epstopdf} 
\DeclareGraphicsExtensions{.pdf,.eps,.png,.jpg,.mps}
\usepackage{hyperref}
\usepackage{bm,amsmath}
\usepackage{dcolumn}

\newcommand{\bra}[1]{\left\langle #1 \right|}
\newcommand{\ket}[1]{\left|#1\right\rangle}
\newcommand{\braket}[2]{\left\langle#1 |  #2\right\rangle}

\def\BEq{\begin{equation}}
\def\EEq{\end{equation}}
\def\BEqA{\begin{eqnarray}}
\def\EEqA{\end{eqnarray}}

\begin{document}

\title{A note on the measures of process fidelity for non-unitary quantum operations}

\author{Joydip Ghosh}
\email{jghosh@physast.uga.edu}

\affiliation{Department of Physics and Astronomy, University of Georgia, Athens, Georgia 30602}

\date{\today}

\begin{abstract}

Various fidelity measures can be defined between two quantum processes especially when at least one of them is non-unitary. In this paper we consider two such measures of state averaged process fidelity, put forward an efficient procedure to compute them and investigate their mathematical properties as well as their relationship.

\end{abstract}


\maketitle

\section{Introduction}
\label{sec:INTRO}

The destruction of classical determinism in the quantum limit necessitates a probabilistic description of the quantum reality. The understanding of quantum states and their evolution in a Hilbert space is an attempt to reconstruct the deterministic nature of quantum mechanics from a different perspective. Hilbert space in quantum mechanics plays the role of phase space in classical mechanics and therefore the question of closeness between two quantum states becomes legitimate \cite{ncbook,PhysRevA.71.062310,springerlink:10.1007/s11128-010-0166-1}. However, for situations where quantum operations are of central importance, such as in quantum computing \cite{ncbook}, an equally relevant question is how to quantify the closeness between two quantum operations. The requirement of defining a measure of fidelity between two quantum processes has so far been considered mostly for cases where the relevant quantum operations are unitary \cite{hill2007prl,Pedersen200747,PhysRevA.81.052340}, which motivates us to investigate a more practical question of introducing fidelity measures in the context of non-unitary quantum operations\cite{Pedersen20087028}.   

The non-unitary quantum operations are relevant in various areas especially in superconducting quantum computing where the quantum gates are implemented within a subspace of the full Hilbert space \cite{StrauchPRL03}. In such cases the leakage outside the computational subspace constitutes an integral part of the gate error. The goal here is to introduce fidelity measures that can quantify this error efficiently. In this paper we consider two such measures. The first one is a natural generalization of the unitary case \cite{Pedersen20087028}, where we project the full time-evolution operator ($U_{f}$) first to the computational subspace (to obtain $U_{s}$) and then apply the fidelity definition for unitary operations between $U_{s}$ and our target quantum gate $U_{\rm tgt}$. In the second measure we embed our unitary target $U_{\rm tgt}$ inside a bigger unitary operator $V_{\rm tgt}$ having the same dimension as $U_{f}$ and then maximize the fidelity (obtained via fidelity definition for unitary case) over the non-computational subspace of $V_{\rm tgt}$. The motivation behind the second definition of fidelity measure stems from the fact that in a real experiment we produce the entire time evolution operator $U_{f}$, but measuring the error only in the computational subspace tacitly assumes that the error in the non-computational subspace is minimum if not zero. This can be achieved only if the fidelity is maximized over various choices of constructing a target in the non-computational subspace of $V_{\rm tgt}$. In this paper we explore the mathematical properties of these two fidelity measures and derive a relationship between them. We show that the expensive computation involved in the maximization of the second definition can be overcome by obtaining the maximum value analytically applying the polar decomposition theorem. Our numerical result reveals quantitative features of these fidelity measures for a realistic situation.

The rest of the paper is organized as follows. In Sec.(\ref{fidelitymeasures}) we first describe the fidelity measure for unitary operations and then introduce two measures of fidelity of non-unitary quantum operations. In Sec.(\ref{analysis}) we analyse various mathematical properties of the fidelity measures. In Sec.(\ref{implication}) we perform an analytical comparison between the two definitions. In Sec.(\ref{nut}) we discuss how our results can be generalized for cases where the target quantum operation is also non-unitary. We demonstrate our analytics with a numerical example in Sec.(\ref{numex}) and discuss the conclusion and scopes of our work in Sec.(\ref{conclusion}).

\section{Fidelity Measures}
\label{fidelitymeasures}

In this section, we discuss fidelity measures for unitary and non-unitary time-evolution operators.

\subsection{Unitary Operation}

Fidelity between two quantum states denotes their distance in the Hilbert space and can be defined as,

\BEq
F(\ket{\psi},\ket{\chi}) \equiv \left|\braket{\psi}{\chi}\right|^{2},
\EEq
where $\ket{\psi}$ and $\ket{\chi}$ are two (pure) states. We can always define a state dependent fidelity between two quantum processes ($U$ and $U_{t}$) as,
\BEq
F_{\chi}(U_{t},U) \equiv \left|\bra{\chi}U_{t}^{\dagger}U\ket{\chi}\right|^{2},
\EEq
where $\ket{\chi}$ is an arbitrary quantum state on which the quantum operations are applied. For cases when $U$ and $U_{t}$ are both unitary, the state averaged process fidelity between them can be defined as,
\BEq\label{favgdef}
F_{\rm avg}(U_{t},U)\equiv\int \left|\bra{\chi}U_{t}^{\dagger}U\ket{\chi}\right|^{2} dV,
\EEq
where $\ket{\chi}$ is a (n dimensional) normalized state vector defined on the unit sphere $S^{2n-1}$ in $\mathbb{C}$ and $dV$ being a normalized measure. It can be shown \cite{Pedersen20087028,Pedersen200747} that such a state averaged fidelity formula can also be expressed as a trace formula given by,
\BEq\label{favgres}
F_{\rm avg}(U_{t},U)=\frac{Tr(U_{t}^{\dagger}UU^{\dagger}U_{t})+\left|Tr(U_{t}^{\dagger}U)\right|^{2}}{n(n+1)},
\EEq
$n$ being the dimensionality of the Hilbert space. While the result in Eq.(\ref{favgres}) being a mathematical identity irrespective of the unitarity of $U$ and $U_{t}$, the definition in Eq.(\ref{favgdef}) is necessarily applicable for situations when both $U_{t}$ and $U$ are unitary. Such cases arise when we compute the fidelity between two quantum operations defined on a full Hilbert space or a subspace which is absolutely decoupled from the rest of the Hilbert space. In quantum computing such cases are highly unlikely and therefore the interesting question is how to define a fidelity measure between two quantum operations where the actual time-evolution operator $U$ may be non-unitary while our target quantum operation $U_{\rm tgt}$ is unitary.

\subsection{Non-unitary Operation}

There exists at least two different ways to define a fidelity measure for non-unitary cases and in this paper we investigate their mathematical properties and the relationship between them. In quantum computing usually the non-unitary cases arise when we have auxiliary states \cite{StrauchPRL03} or coupled subsystems \cite{2011arXiv1105.3997G}. We are interested to compute the fidelity of a quantum operation ($U_{s}$) which lies in a subspace of the full Hilbert space and therefore non-unitary in general. If we denote the full time-evolution operator (unitary) defined on the entire Hilbert space by $U_{f}$, then in some matrix representation, 
\BEq\label{ufdef}
U_{f} \equiv 
\begin{pmatrix}
 U_{s} & | &   & A &  \\
 - & + & - & - & -  \\
   & | &   &   &  \\
 B & | &   & C &  \\
   & | &   &   &  \\
 \end{pmatrix}
\EEq
where $U_{s}$ is $m$ dimensional, $U_{f}$ is $n$ dimensional and $A$, $B$, $C$ are other blocks of the full time-evolution operator ($U_{f}$). The only constraint we require here is $U_{f}$ to be unitary. 

One way to define fidelity is to construct $U_{s}$ from the full time-evolution operator $U_{f}$ projecting it to the relevant computational subspace and then compute the fidelity between $U_s$ and the target quantum operation $U_{\rm tgt}$ as it is defined in Eq.(\ref{favgdef}),
\BEqA\label{f1def}
F_{1}(U_{\rm tgt},U_{s})\equiv\int \left|\bra{\chi}U_{\rm tgt}^{\dagger}U_{s}\ket{\chi}\right|^{2} dV \nonumber \\
=\frac{Tr(U_{s}U_{s}^{\dagger})+\left|Tr(U_{\rm tgt}^{\dagger}U_{s})\right|^{2}}{m(m+1)},
\EEqA
where both $U_{s}$ and $U_{\rm tgt}$ are $m$ dimensional as shown in Eq.(\ref{ufdef}).

Another way to define fidelity is to embed the target quantum operation $U_{\rm tgt}$ ($m$ dimensional)  inside an $n$ dimensional unitary matrix $V_{\rm tgt}$ as,
\BEq\label{vtgtdef}
V_{\rm tgt} \equiv 
\begin{pmatrix}
 U_{\rm tgt} & | &   & J &  \\
 - & + & - & - & -  \\
   & | &   &   &  \\
 K & | &   & L &  \\
   & | &   &   &  \\
 \end{pmatrix},
\EEq 
and then choose $J$, $K$, $L$ such that the fidelity (as defined in Eq.(\ref{favgdef})) between $U_{f}$ and $V_{\rm tgt}$ is maximum. This maximisation is required because in a real experiment the error of the gate operation performed in the non-computational subspace is implicitly assumed to be minimum and therefore it gives an alternative fidelity measure between $U_{\rm tgt}$ and $U_{s}$ and is given by,
\BEqA\label{f2def}
F_{2}(U_{\rm tgt},U_{s}) \equiv F'_{2}(U_{\rm tgt},U_{s}){\vert}_{\rm max\lbrace J,K,L\rbrace}, \nonumber \\
F'_{2}(U_{\rm tgt},U_{s}) \equiv \int \left|\bra{\chi}V_{\rm tgt}^{\dagger}U_{f}\ket{\chi}\right|^{2} dV \nonumber \\
=\frac{Tr(U_{f}U_{f}^{\dagger})+\left|Tr(V_{\rm tgt}^{\dagger}U_{f})\right|^{2}}{n(n+1)}=\frac{n+\left|Tr(V_{\rm tgt}^{\dagger}U_{f})\right|^{2}}{n(n+1)},
\EEqA
where ${\rm max\lbrace J,K,L\rbrace}$ means we choose a set $\lbrace J,K,L\rbrace$ for which $F'_{2}$ is maximum. The last equality holds true as $U_{f}$ is always unitary.

\section{Analysis}\label{analysis}

First, we explore the mathematical properties of these definitions. It's interesting to note that if we require both $U_{\rm tgt}$ and $V_{\rm tgt}$ to be unitary then it's possible only if $J=K=0$ and $L$ is unitary. This enables us to write $V_{\rm tgt}$ as a direct sum of $U_{\rm tgt}$ and $L$ as,
\BEq
V_{\rm tgt}=U_{\rm tgt}\oplus L
\EEq
We can write the $\left|Tr(V_{\rm tgt}^{\dagger}U_{f})\right|$ as,
\BEq
\left|Tr(V_{\rm tgt}^{\dagger}U_{f})\right|=\left|Tr(U_{\rm tgt}^{\dagger}U_{s})+Tr(L^{\dagger}C)\right|.
\EEq
Let's denote,
\BEq
Tr(U_{\rm tgt}^{\dagger}U_{s}) \ {\rightarrowtail} \ re^{i\theta} \ \ {\rm and} \ \ Tr(L^{\dagger}C) \ {\rightarrowtail} \ se^{i\phi}.
\EEq
Now, we can write $F'_{2}$ in terms of $r$, $s$ and $F_{1}$ as,
\BEq\label{f2prime}
F'_{2}=\frac{n+m(m+1)F_{1}-Tr(U_{s}U_{s}^{\dagger})+s^{2}+2rs\cos(\theta-\phi)}{n(n+1)},
\EEq
where $s \equiv \left|Tr(L^{\dagger}C)\right|$. We show that $\left|Tr(L^{\dagger}C)\right|$ must have a finite upper limit using the following theorem.

%

\textit{Theorem 1: If $X$ (not necessarily unitary) be a $p{\times}q$ dimensional block inside a unitary matrix $W$, then $Tr(XX^{\dagger}) \leqslant \min\lbrace p,q\rbrace$.}

\textit{Proof}: Let's write the unitary matrix $W$ as,
\BEq
W \equiv 
\begin{pmatrix}
 X & | &   & Y &  \\
 - & + & - & - & -  \\
   & | &   &   &  \\
  Z & | &  & ...   &  \\
   & | &   &   &  \\
 \end{pmatrix}. \nonumber
\EEq 
Since $W$ is unitary, $Tr(XX^{\dagger})+Tr(YY^{\dagger}) = p$ and also $Tr(X^{\dagger}X)+Tr(Z^{\dagger}Z) = q$. For any matrix $T$, $Tr(TT^{\dagger})$ is a positive real number. Therefore, $Tr(XX^{\dagger}) \leqslant p$ and also $Tr(XX^{\dagger}) \leqslant q$, that altogether gives $Tr(XX^{\dagger}) \leqslant \min\lbrace p,q\rbrace$. Physically the maximal case occurs when the entire population leaks from outside to the blocked subspace.

For our purpose since $L$ is unitary, $Tr(L^{\dagger}L) = n-m$ and using Theorem 1, $Tr(C^{\dagger}C) \leqslant n-m$ and therefore (using Cauchy-Schwarz inequality) we can show,
\BEq
s \equiv \left|Tr(L^{\dagger}C)\right| \leqslant \sqrt{Tr(L^{\dagger}L)Tr(C^{\dagger}C)} \leqslant n-m
\EEq
It is also possible to determine the maximum value of $s$ ($s_{\rm max}$) by using the polar decomposition theorem. According to polar decomposition theorem any square matrix can be written as a product of a unitary and a hermitian matrix, where the hermitian factor is always unique. So we can write \cite{higham:1178} $C$ as, $C = C_{\rm U}C_{\rm H}$ where $C_{\rm H} = \sqrt{C^{\dagger}C}$, the square root here denoting the principal square root of a matrix where all the eigenvalues have a non-negative real part. Since the $C_{\rm U}$ is the closest unitary matrix to $C$ \cite{higham:1178}, we can choose our $L$ to be equal to $C_{\rm U}$ and write,
\BEq\label{smaxformula}
s_{\rm max}=\left| Tr(C_{\rm U}^{\dagger}C_{\rm U}C_{\rm H})\right| = \left| Tr(\sqrt{C^{\dagger}C})\right|,
\EEq
where the square root denotes the principal square root of a matrix.

Now as we can see from Eq.(\ref{f2prime}), in order to maximize $F'_{2}$ we need to satisfy two conditions: i) we need to choose an $L$ for which $s$ is maximum (given by Eq.(\ref{smaxformula})) and ii) we need to choose an $L$ for which $\theta = \phi$. Suppose $L_{\rm max}$ satisfies these conditions and $s_{\rm max}$ being the maximum value of $s$. We can define a fidelity between $C$ and $L$ as,
\BEq\label{foutdef}
F_{\rm out}(L,C) \equiv F_{\rm avg}(L,C) = \frac{Tr(C^{\dagger}C)+\left|Tr(L^{\dagger}C)\right|^{2}}{(n-m)(n-m+1)}.
\EEq
We can see from Eq.(\ref{foutdef}) that $s$ is maximum iff $F_{\rm out}(L,C)$ is maximum, but we can't find a unitary $L$ for which $F_{\rm out}$ is unity if $C$ is non-unitary. In such situations $L_{\rm max}$ actually denotes a unitary matrix closest to non-unitary matrix $C$ (which is $C_{\rm U}$ as defined earlier). If $F^{\rm max}_{\rm out}$ is the maximum possible fidelity between $L$ and $C$, then from Eq.(\ref{smaxformula} and \ref{foutdef}), we can write,
\BEq
F_{\rm out}^{\rm max}(L,C) = \frac{Tr(C^{\dagger}C)+\left|Tr(\sqrt{C^{\dagger}C})\right|^{2}}{(n-m)(n-m+1)}.
\EEq
Now we write our second definition of fidelity in terms of first one as,
\begin{widetext}
\BEq\label{f2res}
F_{2} =\frac{n+\left| Tr(\sqrt{C^{\dagger}C})\right|^{2}+m(m+1)F_{1} -Tr(U_{s}U_{s}^{\dagger})}{n(n+1)} +\frac{2 \left| Tr(\sqrt{C^{\dagger}C})\right|\sqrt{m(m+1)F_{1} -Tr(U_{s}U_{s}^{\dagger})}}{n(n+1)},
\EEq
\end{widetext}
which means $F_{1}$ and $F_{2}$ carries a non-linear relationship between them. 

\section{Implication}\label{implication}

Eq.(\ref{f2res}) gives a neat formula between two definitions of fidelity. In this section we compare these two definitions. First, we show analytically that $F_{2} \geqslant F_{1}$ if $U_{s}$ is unitary. This actually means we need to prove the following inequality,
\BEq\label{f1f2ineq}
\frac{m+\left| Tr(U_{\rm tgt}^{\dagger}U_{s})\right| ^{2}}{m+m^{2}} \leqslant \frac{n+\left( n-m+\left| Tr(U_{\rm tgt}^{\dagger}U_{s})\right| \right)^{2}}{n+n^{2}},
\EEq
since if $U_{s}$ is unitary then so is $C$. Notice that for this case (using Cauchy-Schwarz and Theorem 1), $\left| Tr(U_{\rm tgt}^{\dagger}U_{s})\right| \leqslant m$. Let's denote $\left( m-\left| Tr(U_{\rm tgt}^{\dagger}U_{s})\right|\right)$ by $\delta$ and therefore, $0 \leqslant \delta \leqslant m$. So, in order to show Eq.(\ref{f1f2ineq}), we need to prove the following inequality for any $0 \leqslant \delta \leqslant m$,
\BEq\label{f1f2ineqrefined}
\frac{m+(m-\delta)^{2}}{m+m^{2}} \leqslant \frac{n+\left( n-\delta\right)^{2}}{n+n^{2}}.
\EEq
Let's start with the fact that $n>m$ which implies $n \geqslant m+1$ and after a little manipulation this gives $m \leqslant \frac{2mn}{m+n+1}$. Combining this result with the condition that $0 \leqslant \delta \leqslant m$, we write,  $\delta \leqslant \frac{2mn}{m+n+1}$. An algebraic manipulation over this step actually gives us $\frac{2m-\delta}{m+m^{2}} \geqslant \frac{2n-\delta}{n+n^{2}}$ which ultimately gives rise to Eq.(\ref{f1f2ineqrefined}) and thus we can prove $F_2 \geqslant F_{1}$ when $U_{s}$ is unitary.

Now in order for everything to make sense we need to show that at higher values of $F_{1}$ there should not be much difference between $F_{1}$ and $F_{2}$ and we show here that at $\lim_{F_{1} \rightarrow 1} F_{2} = 1$. At $\lim_{F_{1} \rightarrow 1}$, $U_{s}$ tends to be a unitary matrix and so is $C$ and therefore plugging everything in Eq.(\ref{f2res}), we obtain,
\begin{align*}
\lim_{F_{1} \rightarrow 1} F_{2} &= \frac{n+(n-m)^{2}+m(m+1)-m+2(n-m)m}{n(n+1)} \nonumber \\
 &= 1.
\end{align*}
So, both the fidelity measures are identical at unity.

\section{Non-unitary Target}\label{nut}
It is possible to generalize our analysis for non-unitary target operations. Such a situation is unlikely in the context of quantum computing (without decoherence), but may arise in the context of quantum simulation. By non-unitary target operation we actually mean our $U_{\rm tgt}$ to be non-unitary while we still require $V_{\rm tgt}$ and $U_{f}$ to be always unitary and as we can note for such cases $J \neq K \neq 0$ in general. This essentially means that we need to replace $L^{\dagger}C$ by $J^{\dagger}A+K^{\dagger}B+L^{\dagger}C$ in our previous analysis which gives,
\BEq
F'_{2}=\frac{n+m(m+1)F_{1}-Tr(U_{s}U_{s}^{\dagger})+s^{2}+2rs\cos(\theta-\phi)}{n(n+1)},
\EEq
where, 
\BEq
Tr(U_{\rm tgt}^{\dagger}U_{s}) \ {\rightarrowtail} \ re^{i\theta} \ \ {\rm and} \ \ Tr(J^{\dagger}A+K^{\dagger}B+L^{\dagger}C) \ {\rightarrowtail} \ se^{i\phi}
\EEq
for this case. In order for $F'_{2}$ to be maximum we need to choose $J$, $K$ and $L$ such that $s$ is maximum. So, it's important to show that such a maxima exists. Applying triangle inequality we write,
\BEq
s \leqslant \left| Tr(J^{\dagger}A)\right| + \left| Tr(K^{\dagger}B)\right| +\left| Tr(L^{\dagger}C) \right|.
\EEq
And finally using Cauchy-Schwarz inequality and our Theorem 1, we can show that $s \leqslant m+n$; $m$, $n$ being dimensions of $U_{\rm tgt}$ and $V_{\rm tgt}$ respectively. This actually means that $s$ must have a finite upper bound and suppose $s_{\rm max}$ be its possible maximum value. Then we can write,
\BEq
F_{2}=\frac{n+m(m+1)F_{1}-Tr(U_{s}U_{s}^{\dagger})+s_{\rm max}^{2}+2rs_{\rm max}}{n(n+1)},
\EEq
which is the relation between two fidelity measures for non-unitary target operations.

\section{Numerical Analysis}\label{numex}

In this section we investigate the two measures of fidelity more quantitatively in the light of a real physical system. We consider a controlled-Z gate implementation in a superconducting phase qubit capacitively coupled to a resonator. The Hamiltonian for this system is given by,

\BEq\label{hamiltonian}
H=H_{0}+H_{\rm int},
\EEq
where,
\BEq\label{hamiltonianParameters}
\begin{array}{l}
H_{0}=\left(\begin{array}{ccc}
0 & 0 & 0 \\
0 & \omega_{\rm q} & 0 \\
0 & 0 & 2\omega_{\rm q}-\Delta\end{array}\right)_{q}  + \left(\begin{array}{ccc}
0 & 0 & 0 \\
0 & \omega_{\rm r} & 0 \\
0 & 0 & 2\omega_{\rm r}\end{array}\right)_{r}\\
\\
H_{\rm int}=g \left(\begin{array}{ccc}
0 & -i & 0 \\
i & 0 & -\sqrt{2}i \\
0 & \sqrt{2}i & 0\end{array}\right)_{q}\otimes \left(\begin{array}{ccc}
0 & -i & 0 \\
i & 0 & -\sqrt{2}i \\
0 & \sqrt{2}i & 0\end{array}\right)_{r},
\end{array}
\EEq
where, $\omega_{\rm q}$, $\omega_{\rm r}$ and $\Delta$ are qubit frequency, resonator frequency, and anharmonicity of the qubit, respectively and $g$ is the (time-independent) interaction strength between the qubit and resonator. The Hamiltonian used here has nine levels in it while four of those constitute a computational subspace. So we can define and compute both the fidelities for this situation. The purpose here is to compare two fidelity measures quantitatively and in order to simplify our problem we set the qubit frequency $\omega_{\rm q}$ (6.2 GHz.) at $\omega_{\rm r}+\Delta$, resonator frequency $\omega_{\rm r}$ at 6 GHz. and coupling ($g$) at 30 MHz. and consider our Hamiltonian to be time-independent.

\begin{figure}[htb]
	\centering
\includegraphics[angle=0,width=1.00\linewidth]{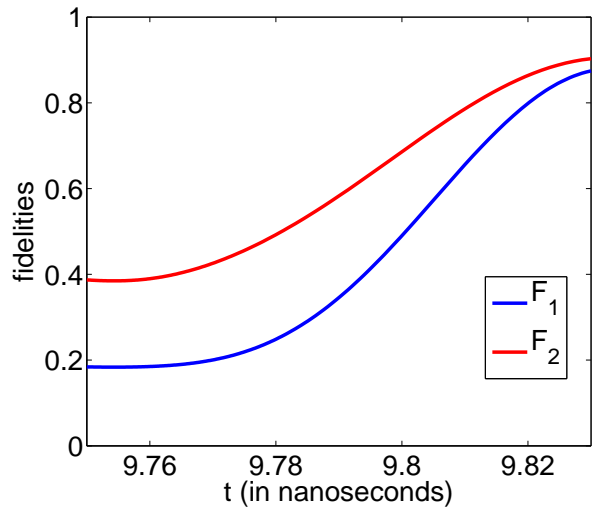}
				\caption{(Color online) A plot of two fidelity measures vs. time (in nanoseconds) where the target operation is a controlled-Z gate in a resonator coupled phase qubit.}
		\label{fig:fidelities}
\end{figure}

We consider the time evolution of our Hamiltonian for a short span near the regime where we expect to obtain a controlled-Z gate (Strauch's approach \cite{StrauchPRL03}) and plot both the fidelities (Fig.(\ref{fig:fidelities})) with respect to time. Our numerical result shows that $F_{2} > F_{1}$ for that regime. We have proved analytically that $F_{2} > F_{1}$ if $U_{s}$ is unitary, while proving (or disproving) the result $F_{2} \geqslant F_{1}$ universally (for non-unitary situation) has not been accomplished in our work and therefore is left as an open problem. 

\section{Conclusion}\label{conclusion}
In summary, we have considered two possible measures of fidelity for non-unitary quantum operations and explored their mathematical properties. A relation between them has been established. It is found that the maximization involved in the second definition of fidelity can be done analytically and the optimal value is expressed in closed form. Our numerical result shows the quantitative feature of the two fidelities for a realistic situation. The plot in Fig.(\ref{fig:fidelities}) shows that while the two measures of fidelity differ considerably for lower values, they converge for higher values. The difference between two measures of fidelities at lower values is not surprising and shows up even for unitary case. For example, if we look into the relationship between the fidelity measures (for unitary situation) considered by Hill (let's call it $F_{H}$. See Eq.(13) in Ref.\cite{hill2007prl}.) and the one considered by Pedersen \textit{et. al.} (Let's call it $F_{P}$. See Ref.\cite{Pedersen200747}), we can write,
\BEq\label{fhfp}
F_{P} = \frac{1+nF^{2}_{H}}{1+n},
\EEq
$n$ being the dimension of the Hilbert space. Eq.(\ref{fhfp}) tells us that at $F_{H} \rightarrowtail 1$, $F_{P} \rightarrowtail 1$, but at $F_{H} \rightarrowtail 0$, $F_{P} \rightarrowtail \frac{1}{1+n}$. The difference between fidelity measures for lower fidelities resemble the fact that in statistics different measures of central tendency (mean, median and mode) differ a lot in general, while give identical values when all numbers the data set are equal. Our numerical results also show that $F_{2} \geqslant F_{1}$. The proof (or contradiction) of this result for non-unitary cases is still lacking and can be considered as a future direction of research on this subject. Our result only shows the relationship between two possible measures of state averaged process fidelity when a non-unitary time-evolution operator is produced in an experiment. Similar questions can be investigated as a follow-up of this work for state-minimized (worst case) process fidelity which is more relevant to quantum error correction.

\begin{acknowledgments}
It is a pleasure to thank Andrei Galiautdinov, Michael Geller, Sayonita Ghoshhajra, Emily Pritchett and Zhongyuan Zhou for useful discussions.
\end{acknowledgments}

\bibliography{fidelity}

\begin{thebibliography}{10}%
\makeatletter
\providecommand \@ifxundefined [1]{%
 \@ifx{#1\undefined}
}%
\providecommand \@ifnum [1]{%
 \ifnum #1\expandafter \@firstoftwo
 \else \expandafter \@secondoftwo
 \fi
}%
\providecommand \@ifx [1]{%
 \ifx #1\expandafter \@firstoftwo
 \else \expandafter \@secondoftwo
 \fi
}%
\providecommand \natexlab [1]{#1}%
\providecommand \enquote  [1]{``#1''}%
\providecommand \bibnamefont  [1]{#1}%
\providecommand \bibfnamefont [1]{#1}%
\providecommand \citenamefont [1]{#1}%
\providecommand \href@noop [0]{\@secondoftwo}%
\providecommand \href [0]{\begingroup \@sanitize@url \@href}%
\providecommand \@href[1]{\@@startlink{#1}\@@href}%
\providecommand \@@href[1]{\endgroup#1\@@endlink}%
\providecommand \@sanitize@url [0]{\catcode `\\12\catcode `\$12\catcode
  `\&12\catcode `\#12\catcode `\^12\catcode `\_12\catcode `\%12\relax}%
\providecommand \@@startlink[1]{}%
\providecommand \@@endlink[0]{}%
\providecommand \url  [0]{\begingroup\@sanitize@url \@url }%
\providecommand \@url [1]{\endgroup\@href {#1}{\urlprefix }}%
\providecommand \urlprefix  [0]{URL }%
\providecommand \Eprint [0]{\href }%
\@ifxundefined \urlstyle {%
  \providecommand \doi  [0]{\begingroup \@sanitize@url \@doi}%
  \providecommand \@doi [1]{\endgroup \@@startlink {\doibase
  #1}doi:\discretionary {}{}{}#1\@@endlink }%
}{%
  \providecommand \doi  [0]{doi:\discretionary{}{}{}\begingroup
  \urlstyle{rm}\Url }%
}%
\providecommand \doibase [0]{http://dx.doi.org/}%
\providecommand \Doi [0]{\begingroup \@sanitize@url \@Doi }%
\providecommand \@Doi  [1]{\endgroup\@@startlink{\doibase#1}\@@Doi}%
\providecommand \@@Doi [1]{#1\@@endlink}%
\providecommand \selectlanguage [0]{\@gobble}%
\providecommand \bibinfo  [0]{\@secondoftwo}%
\providecommand \bibfield  [0]{\@secondoftwo}%
\providecommand \translation [1]{[#1]}%
\providecommand \BibitemOpen [0]{}%
\providecommand \bibitemStop [0]{}%
\providecommand \bibitemNoStop [0]{.\EOS\space}%
\providecommand \EOS [0]{\spacefactor3000\relax}%
\providecommand \BibitemShut  [1]{\csname bibitem#1\endcsname}%
\bibitem [{\citenamefont {Nielsen}\ and\ \citenamefont
  {Chuang}(2005)}]{ncbook}%
  \BibitemOpen
  \bibfield  {author} {\bibinfo {author} {\bibfnamefont {M.~A.}\ \bibnamefont
  {Nielsen}}\ and\ \bibinfo {author} {\bibfnamefont {I.~L.}\ \bibnamefont
  {Chuang}},\ }\href@noop {} {\emph {\bibinfo {title} {Quantum Computation and
  Quantum Information}}}\ (\bibinfo  {publisher} {Cambridge University Press},\
  \bibinfo {year} {2005})\BibitemShut {NoStop}%
\bibitem [{\citenamefont {Gilchrist}\ \emph {et~al.}(2005)\citenamefont
  {Gilchrist}, \citenamefont {Langford},\ and\ \citenamefont
  {Nielsen}}]{PhysRevA.71.062310}%
  \BibitemOpen
  \bibfield  {author} {\bibinfo {author} {\bibfnamefont {A.}~\bibnamefont
  {Gilchrist}}, \bibinfo {author} {\bibfnamefont {N.~K.}\ \bibnamefont
  {Langford}}, \ and\ \bibinfo {author} {\bibfnamefont {M.~A.}\ \bibnamefont
  {Nielsen}},\ }\Doi {10.1103/PhysRevA.71.062310} {\bibfield  {journal}
  {\bibinfo  {journal} {Phys. Rev. A},\ }\textbf {\bibinfo {volume} {71}},\
  \bibinfo {pages} {062310} (\bibinfo {year} {2005})}\BibitemShut {NoStop}%
\bibitem [{\citenamefont {Puchaùa}\ \emph {et~al.}(2011)\citenamefont
  {Puchaùa}, \citenamefont {Miszczak}, \citenamefont {Gawron},\ and\
  \citenamefont {Gardas}}]{springerlink:10.1007/s11128-010-0166-1}%
  \BibitemOpen
  \bibfield  {author} {\bibinfo {author} {\bibfnamefont {Z.}~\bibnamefont
  {Puchaùa}}, \bibinfo {author} {\bibfnamefont {J.}~\bibnamefont {Miszczak}},
  \bibinfo {author} {\bibfnamefont {P.}~\bibnamefont {Gawron}}, \ and\ \bibinfo
  {author} {\bibfnamefont {B.}~\bibnamefont {Gardas}},\ }\href
  {http://dx.doi.org/10.1007/s11128-010-0166-1} {\bibfield  {journal} {\bibinfo
   {journal} {Quantum Information Processing},\ }\textbf {\bibinfo {volume}
  {10}},\ \bibinfo {pages} {1} (\bibinfo {year} {2011})},\ ISSN \bibinfo {issn}
  {1570-0755},\ \bibinfo {note} {10.1007/s11128-010-0166-1}\BibitemShut
  {NoStop}%
\bibitem [{\citenamefont {Hill}(2007)}]{hill2007prl}%
  \BibitemOpen
  \bibfield  {author} {\bibinfo {author} {\bibfnamefont {C.~D.}\ \bibnamefont
  {Hill}},\ }\Doi {10.1103/PhysRevLett.98.180501} {\bibfield  {journal}
  {\bibinfo  {journal} {Phys. Rev. Lett.},\ }\textbf {\bibinfo {volume} {98}},\
  \bibinfo {eid} {180501} (\bibinfo {year} {2007})}\BibitemShut {NoStop}%
\bibitem [{\citenamefont {Pedersen}\ \emph {et~al.}(2007)\citenamefont
  {Pedersen}, \citenamefont {M{\o}ller},\ and\ \citenamefont
  {M{\o}lmer}}]{Pedersen200747}%
  \BibitemOpen
  \bibfield  {author} {\bibinfo {author} {\bibfnamefont {L.~H.}\ \bibnamefont
  {Pedersen}}, \bibinfo {author} {\bibfnamefont {N.~M.}\ \bibnamefont
  {M{\o}ller}}, \ and\ \bibinfo {author} {\bibfnamefont {K.}~\bibnamefont
  {M{\o}lmer}},\ }\Doi {DOI: 10.1016/j.physleta.2007.02.069} {\bibfield
  {journal} {\bibinfo  {journal} {Physics Letters A},\ }\textbf {\bibinfo
  {volume} {367}},\ \bibinfo {pages} {47 } (\bibinfo {year} {2007})},\ ISSN
  \bibinfo {issn} {0375-9601}\BibitemShut {NoStop}%
\bibitem [{\citenamefont {Ghosh}\ and\ \citenamefont
  {Geller}(2010)}]{PhysRevA.81.052340}%
  \BibitemOpen
  \bibfield  {author} {\bibinfo {author} {\bibfnamefont {J.}~\bibnamefont
  {Ghosh}}\ and\ \bibinfo {author} {\bibfnamefont {M.~R.}\ \bibnamefont
  {Geller}},\ }\Doi {10.1103/PhysRevA.81.052340} {\bibfield  {journal}
  {\bibinfo  {journal} {Phys. Rev. A},\ }\textbf {\bibinfo {volume} {81}},\
  \bibinfo {pages} {052340} (\bibinfo {year} {2010})}\BibitemShut {NoStop}%
\bibitem [{\citenamefont {Pedersen}\ \emph {et~al.}(2008)\citenamefont
  {Pedersen}, \citenamefont {M{\o}ller},\ and\ \citenamefont
  {M{\o}lmer}}]{Pedersen20087028}%
  \BibitemOpen
  \bibfield  {author} {\bibinfo {author} {\bibfnamefont {L.~H.}\ \bibnamefont
  {Pedersen}}, \bibinfo {author} {\bibfnamefont {N.~M.}\ \bibnamefont
  {M{\o}ller}}, \ and\ \bibinfo {author} {\bibfnamefont {K.}~\bibnamefont
  {M{\o}lmer}},\ }\Doi {DOI: 10.1016/j.physleta.2008.10.034} {\bibfield
  {journal} {\bibinfo  {journal} {Physics Letters A},\ }\textbf {\bibinfo
  {volume} {372}},\ \bibinfo {pages} {7028 } (\bibinfo {year} {2008})},\ ISSN
  \bibinfo {issn} {0375-9601}\BibitemShut {NoStop}%
\bibitem [{\citenamefont {Strauch}\ \emph {et~al.}(2003)\citenamefont
  {Strauch}, \citenamefont {Johnson}, \citenamefont {Dragt}, \citenamefont
  {Lobb}, \citenamefont {Anderson},\ and\ \citenamefont
  {Wellstood}}]{StrauchPRL03}%
  \BibitemOpen
  \bibfield  {author} {\bibinfo {author} {\bibfnamefont {F.~W.}\ \bibnamefont
  {Strauch}}, \bibinfo {author} {\bibfnamefont {P.~R.}\ \bibnamefont
  {Johnson}}, \bibinfo {author} {\bibfnamefont {A.~J.}\ \bibnamefont {Dragt}},
  \bibinfo {author} {\bibfnamefont {C.~J.}\ \bibnamefont {Lobb}}, \bibinfo
  {author} {\bibfnamefont {J.~R.}\ \bibnamefont {Anderson}}, \ and\ \bibinfo
  {author} {\bibfnamefont {F.~C.}\ \bibnamefont {Wellstood}},\ }\href@noop {}
  {\bibfield  {journal} {\bibinfo  {journal} {Phys. Rev. Lett.},\ }\textbf
  {\bibinfo {volume} {91}},\ \bibinfo {pages} {167005} (\bibinfo {year}
  {2003})}\BibitemShut {NoStop}%
\bibitem [{\citenamefont {Galiautdinov}\ \emph {et~al.}(2011)\citenamefont
  {Galiautdinov}, \citenamefont {Korotkov},\ and\ \citenamefont
  {Martinis}}]{2011arXiv1105.3997G}%
  \BibitemOpen
  \bibfield  {author} {\bibinfo {author} {\bibfnamefont {A.}~\bibnamefont
  {Galiautdinov}}, \bibinfo {author} {\bibfnamefont {A.~N.}\ \bibnamefont
  {Korotkov}}, \ and\ \bibinfo {author} {\bibfnamefont {J.~M.}\ \bibnamefont
  {Martinis}},\ }\href@noop {} {\bibfield  {journal} {\bibinfo  {journal}
  {ArXiv e-prints}} (\bibinfo {year} {2011})},\ \Eprint
  {http://arxiv.org/abs/1105.3997} {arXiv:1105.3997 [quant-ph]} \BibitemShut
  {NoStop}%
\bibitem [{\citenamefont {Higham}\ \emph {et~al.}(2004)\citenamefont {Higham},
  \citenamefont {Mackey}, \citenamefont {Mackey},\ and\ \citenamefont
  {Tisseur}}]{higham:1178}%
  \BibitemOpen
  \bibfield  {author} {\bibinfo {author} {\bibfnamefont {N.~J.}\ \bibnamefont
  {Higham}}, \bibinfo {author} {\bibfnamefont {D.~S.}\ \bibnamefont {Mackey}},
  \bibinfo {author} {\bibfnamefont {N.}~\bibnamefont {Mackey}}, \ and\ \bibinfo
  {author} {\bibfnamefont {F.}~\bibnamefont {Tisseur}},\ }\Doi
  {10.1137/S0895479803426644} {\bibfield  {journal} {\bibinfo  {journal} {SIAM
  Journal on Matrix Analysis and Applications},\ }\textbf {\bibinfo {volume}
  {25}},\ \bibinfo {pages} {1178} (\bibinfo {year} {2004})}\BibitemShut
  {NoStop}%
\end{thebibliography}%

\end{document}